\documentclass[amssymb,preprint,prb,aps]{revtex4}

\begin{document}

{\bf Title}: Statistical properties of acoustic emission signals from metal
cutting processes

F.A. Farrelly, A. Petri, L. Pitolli, G. Pontuale $^{a)}$

\begin{quotation}
Consiglio Nazionale delle Ricerche, Istituto di Acustica ''O.M.Corbino''

Via del Fosso del Cavaliere, 100 - 00133 Roma, Italy
\end{quotation}

A. Tagliani, P.L. Novi Inverardi

\begin{quote}
Faculty of Economics, Trento University

Via Vigilio Inama, 5 - 38100 Trento, Italy
\end{quote}

Received.

\smallskip Running title: Acoustic emission statistical properties

Abbreviated title: Acoustic emission statistical properties

$^{a)}$ Electronic-mail: pontuale@idac.rm.cnr.it

$^{b)}$ Portions of this work were published in ''A statistical analysis of
acoustic emission signals for tool condition monitoring (TCM)'',
ARLO-Acoustic Research Letters Online, 4(1), January 2003.

\smallskip

\pagebreak

{\bf ABSTRACT}

Acoustic Emission (AE) data from single point turning machining are analysed
in this paper in order to gain a greater insight of the signal statistical
properties for Tool Condition Monitoring (TCM) applications. A statistical
analysis of the time series data amplitude and root mean square (RMS) value
at various tool wear levels are performed, finding that ageing features can
be revealed in all cases from the observed experimental histograms. In
particular, AE data amplitudes are shown to be distributed with a power-law
behaviour above a cross-over value. An analytic model for the RMS values
probability density function ({\it pdf}) is obtained resorting to the
Jaynes' maximum entropy principle (MEp); novel technique of constraining the
modelling function under few fractional moments, instead of a greater amount
of ordinary moments, leads to well-tailored functions for experimental
histograms.

\medskip \medskip \medskip \smallskip

\smallskip

PACS numbers: 43.60.Cg, 43.40.Le, 02.60.Gf

\pagebreak

{\bf I. INTRODUCTION}

\smallskip

\smallskip Due to global competition and rapidly changing customer
requirements, enterprises are required to constantly redesign their products
and continuously reconfigure their manufacturing processes in terms of
increasing flexibility and complexity, in order to satisfy the international
market's demands to reduce production costs and increase precision and
quality. Design and development of on-line systems for monitoring the
process parameters, parts and manufacturing environment, is becoming more
and more important, as the actual Sixth Framework European Programme FP6
efforts demonstrate. In this framework, main problems in the field of metal
cutting are constituted by tool wear and tool breakage. These phenomena
limit the cutting speed and feed rate, and consequently, the metal removal
rates that can be used in machining various workpiece materials in an
economic way. Also, this fact plays a negative role in the machine tool
environment causing unexpected breakdowns, defective workpieces, overloads
due to high cutting forces and machine tool damages, as well as other
problems that reduce the productiveness of the machine tool. Usually, these
problems are solved using a conservative limit for the tool useful life,
this leading to a less optimum use of the tool. The complexity of such a
problem has lead to an impressive amount of literature on this subject, and
a variety of techniques have been proposed. An extended review of the state
of the art, technological challenges and future developments of these
systems is described by Byrne et al.~\cite{Byrne}. This paper deals in great
detail on describing the physical parameters to be analysed for industrial
control applications, together with their appropriate sensory systems. Among
these, Acoustic Emission (AE) signal analysis has been demonstrated to be
one of the most efficient TCM techniques which can be applied to machining
processes control, as the impressive amount of literature on this subject
shows; Xiaoli's article \cite{xiaoli} is just an example of brief review
about AE methods for tool wear monitoring during turning machining. Also AE
source identification and modelling, for this particular application, is a
subject in which, during the last years, a large number of studies have been
conducted, (see only as a few important examples \cite{hatano}$^{-}$\cite
{chung-kannatey} ); Heiple et al. \cite{heiple-carpenter} found that the
primary source of AE from single point machining is the sliding friction
between the nose and the flank of the tool and the machined surface. This
kind of friction is related in a complex manner with tool wear and the
material being machined; therefore, depending on machining conditions, the
RMS levels and other AE related values may increase or decrease as the tool
wears, affecting the parameters of the experimental frequency distributions.

In this framework, our paper tackles the problem of gaining greater insight
of the basic statistical properties of AE signals, whose better and deeper
knowledge, besides shedding light upon this fundamental aspect of AE for
this application, may greatly facilitate an appropriate implementation of AE
sensor-based devices leading to efficient TCM systems. To do this,
single-point turning machining conditions, that will be described in the
next section, were held fixed throughout the experiment, so as to limit the
number of varying parameters that might affect the behaviour of the observed
quantities. The experimental probability density functions ({\it pdf}) of AE
time series amplitude and Root Mean Squared (RMS) values are shown for
different levels of tool wear, both these approaches being capable of
showing interesting and not yet completely exploited features. Furthermore,
the effects of tool wear on such statistical properties are highlighted,
thus outlining possible further signal analysis scenarios.

An analytic model for the RMS {\it pdf} reconstruction is presented here,
resorting to the Jaynes' maximum entropy principle (MEp) principle; the
novel technique, recently proposed by some of the authors, of constraining
the modelling function under some fractional moments instead of a greater
amount of ordinary integer moments, leads to well-tailored functions for the
experimental {\it pdf}. These results are compared with previously
considered models, showing a substantial improvement in the agreement with
experimental histograms.

\smallskip

{\bf II. DETECTORS AND EXPERIMENTAL SET-UP}

\smallskip

To achieve the objectives of this work, simultaneous AE data acquisition has
been conducted by means of two different AE sensors: a custom-built AE
sensor, and a Br\"{u}el \& Kj\ae r 8312 AE sensor. The choice of using two
different transducers for signal pick-up not only allows a more reliable and
intensive harvest of data, but also makes it possible to perform a compared
analysis on signals gathered at the same time but at different locations and
in different conditions. In fact, the propagation of AE signals in the range
investigated is characterised by significant attenuation. Thus, in order to
achieve a good signal to noise ratio, the sensor should be placed as close
as possible to the machining point where the AE signal is generated \cite
{Jemielniak1}; as an added benefit, reduction of the signal distortion due
to the number of interfaces and mechanical resonances is also achieved by
avoiding a long measurement chain. This motivated the use of a custom-built
sensor, made of a small rectangular shaped piezoelectric ceramic (PZT-5), $%
5.0\ $x $1.5$ x $0.7~$mm in size, working as a resonant sensor with a
resonance frequency near 370 kHz, housed inside a small cavity bored into
the cutting tool holder so as to protect it from chip damages and liquid
coolant effects, and placed about two centimetres from the AE signal
sources. An electrically conductive adhesive is used to bond the ceramic to
the internal face of the cavity. The commercial sensor is a 40 dB
pre-amplified Br\"{u}el \& Kjaer Type 8312 AE transducer, placed at the
extremity of the tool holder by means of a special mounting, about 12 cm
from the cutting area.

AE\ measurements were performed while machining stainless steel (AISI 303)
bars on a SAG14 GRAZIANO lathe. Cutting speeds range from $0.5$ to $1~m/s$,
while feed rates and cutting depths are kept constant at $0.0195~$mm/turn
and $2~$mm$~$respectively. In all measurements, cutting tool inserts were
''IMPERO'' PCLNR with 2020/12 type tungsten carbide; the acquisitions were
performed on inserts with various degrees of wear. Specifically, inserts
were grouped into three different wear categories: new ones, those estimated
to be half-way through their life-cycle (50\%) and those completely worn
through (100\%).

In the new and 100\% worn cases, 8 cutting edges were analyzed per wear
level, while 4 edges were utilised in the 50\% case. For each edge one
acquisition run was conducted, collecting 15 banks of $40,960$ AE time
series point corresponding to $16.38$ ms, for a total of $614,400$ points
each run. Hence, a total of $12,288,000$ $($ $4.9152$ s$)$ AE time series
points were collected over all 20 runs.

The experimental set-up is roughly sketched in Fig. 1. The signals detected
by the transducers were amplified (by means of a 40 dB Analog Module
preamplifier for the custom sensor, its own 40 dB preamplifier for the
Br\"{u}el \& Kjaer one), and filtered in the $200~-1000~$kHz range through a
Krohn-Hite 3944 filter. The signals were then captured by a Tektronix
digital oscilloscope (TDS420) using a $2.5~$MHz sampling rate, and finally
stored in a PC through an IEEE488 interface. Blank measurements performed
just prior to machining indicated no significant electrical noise. The data
were analysed both directly in their time series form and through Root Mean
Squared (RMS) values.

{\bf III. EXPERIMENTAL RESULTS AND DISCUSSION}

{\bf A. Time series analysis}

\smallskip

Typical time splice series for the two sensors are shown in Fig. 2. In both
cases two rather well distinct parts can be identified: a {\it continuous
part} that is characterised by a relatively constant amplitude with small
fluctuations, and a {\it burst emission} exhibiting strong intermittence and
relatively high amplitudes. The former is associated with plastic
deformation and frictional processes during the cutting operations, the
latest with chip breakage as well with micro-cracks and dislocation kinetics 
\cite{liang-dornfeld }$^{,}$\cite{hatano}.

For the two sensors, the histograms of the absolute value of time series
amplitudes, {\it a}, taken from measurements performed using inserts in
three stages of wear are portrayed in Fig. 3. All these experimental
frequency distributions {\it p(a)} are normalised over the related number of
data and grouped into 126 classes. It is possible to observe how in all
cases the curves exhibit a power-law behaviour $p(a)=Aa^{-\alpha }+B\ $above
a cross-over value from a nearly flat distribution, the value of the slope
being slightly dependent on the sensor used ($\alpha =-3.7$ and $\alpha
=-3.9 $ for custom-built and Br\"{u}el \& Kjaer sensors, respectively), but
similar for all three stages of wear. The corresponding exponents for the
energy {\it E} are $\alpha ^{\prime }=-2.35$ and $\alpha ^{\prime }=-2.45$,
as they can be derived from the amplitude exponents assuming $E\varpropto
a^{2}$.

For both sensors, data from tools with greater wear level show within the
power-law range a slightly smaller frequency count for a given value in
amplitude; this leads to the conclusion that, in this set of trials, the
newer tools are the most active ones in terms of acoustic emission.

It is interesting to note that power-law behaviour, strongly suggestive of a
critical dynamics associated with this particular AE phenomena, has been
observed in many studies on acoustic emission signals, e.g. those related
with the formation of micro-fractures \cite{Petri}$^{,}$\cite{ciliberto}$%
^{,} $\cite{ caldarelli(prl)}. In general, power-law characteristics are
associated with scale invariant properties underlying the physical phenomena
under study, and in some cases this has been explained by Self-Organised
Criticality (SOC) \cite{SOC} models.

\smallskip {\bf B. Root mean squared analysis}

A substantial effort in the past has been dedicated towards analysing the
relationship between signal RMS and tool wear level in various experimental
situations, e.g. see \cite{Jemielniak et al- 1998} for identifying
catastrophic tool failure (CTF) conditions in carbide inserts. The analysis
of the RMS were conducted calculating values on the basis of 100 points,
corresponding to $40$ ms, this choice being effective in making the RMS
signal sensitive to the different contributions from burst and continuous
events. In order to study the RMS values statistical properties, also as a
function of ageing, their experimental frequency distributions were analysed
by grouping the values into 60 bins, after their normalisation over the
largest values of the entire RMS data set. For each wear level, and for both
the sensors utilised, the average histograms are shown in Fig. 4. For
increasing levels of wear the curves show a noticeable shift towards lower
levels of the modal value of the frequency distribution, as well a change in
the skewness tending towards values compatible with a symmetrical shape,
these features being particularly evident for Br\"{u}el \& Kjaer sensor. In
order to test the difference among these graphs, T-Test analyses regarding
the sample means were performed, which indicate that the null hypothesis of
equal means can be rejected with a confidence level of 95\%. This approach
appears to be effective in discriminating tool wear features, and could be
used as the basis for implementing algorithms for TCM applications.

In literature, borrowing from a technique used in the description of
surfaces roughness by Whitehouse \cite{Whitehouse-1978}, various attempts
have been made at determining tool condition relying on the hypothesis that
a Beta distribution $f(x)$ (see for example \cite{Kannatey-Asibu et al- 1982}%
$^{,}$\cite{Jemielniak et al- 1998}) properly describes the Probability
Density Function {\it pdf} of the RMS values,

\begin{equation}
f(x)=\frac{x^{r-1}(1-x)^{s-1}}{\beta (r,s)},  \label{beta}
\end{equation}
where $\beta $ is the complete Beta function: 
\begin{equation}
\beta (r,s)=\int_{0}^{1}x^{r-1}(1-x)^{s-1}dx.  \label{beta_integral}
\end{equation}
With this assumption it is possible to characterize the moments of the
distribution in terms of the two parameters $r$ and $s$, and vice-versa. In
particular, as far as mean $(\mu )$ and variance $(\sigma ^{2})~$are
concerned, we have: 
\begin{eqnarray}
r &=&\frac{\mu }{\sigma ^{2}}(\mu -\mu ^{2}-\sigma ^{2})  \label{rs-stim} \\
s &=&\frac{1-\mu }{\sigma ^{2}}(\mu -\mu ^{2}-\sigma ^{2}).  \nonumber
\end{eqnarray}

Thus, values for {\it r} and {\it s} can be estimated on the basis of the
mean and variance of the data set. Past studies have shown that {\it r,s}
pairs are scattered in different ways, depending on tool conditions \cite
{Kannatey-Asibu et al- 1982}. One shortcoming of this method is that no
estimate of the errors on the {\it r} and {\it s} parameters is directly
available; this is particularly serious as real-life signals often contain
outliers which can bring a noticeable shift in the actual values of both
mean and variance. One possibility is to use more robust estimators (e.g.
median instead of mean) although this still does not give an error estimate
for the calculated parameters. A further choice is to perform a non-linear
best-fit on the data set using the function given in Eq. (1)\cite{footnote}.

In Fig. 5 the best-fit of the experimental frequency distributions from
custom-built sensor data as in Fig. 4 are shown. From these graphs it is
possible to see that while there is a good matching between the fitting
function and the data sets in the neighbourhood of the peaks, some
discrepancies are visible in the residual for RMS bin values just above the
peak where the curves level off; this indicates that in this range, the data
sets are richer in events than what Eq. (1) would indicate, and this
suggests that a better empirical fitting-function may exist. In Fig. 6 {\it %
r,s} estimates from Eqs. (3) are compared to the ones obtained by the
best-fitting process. It is evident that the two groups greatly differ and
that these discrepancies are not compatible considering the error estimates
given on the fitted parameters. Furthermore, the scattering pattern of these
two groups are entirely different; whereas both the best-fitted r,s
parameters tend to increase with wear, the estimated ones show an
essentially opposite behavior. One possible explanation for this difference
is that while the best-fit process minimises mean-square differences between
the fitting function and the frequency distribution (so that heavily
populated bins are weighted more), the estimate method relies on $\mu $ and $%
\sigma ^{2}$. Variance, in particular, is highly sensitive to outliers, so
values far from the mean weigh heavily on its determination.

In this framework, a method is proposed here to reconstruct the approximate
RMS's {\it pdf} by applying the ME technique, under the constraint of some
fractional moments, the latter ones being explicitly obtained in terms of
given ordinary moments. Such approach allows to obtain well-tailored fitting
functions for the experimental curves.

\smallskip

{\bf C. Recovering RMS's {\it pdf} from fractional moments}

Jaynes' maximum entropy principle (MEp) says that {\it ''the best (minimally
prejudiced) assignement of probabilites is that one which minimises the
entropy subject to the satisfaction of the constraints imposed by the
available information''}\cite{jaynes}. Thus, taking the Kullback-Leibler
information functional or differential entropy (KL, in the following) as the
relevant information measure, the spirit of Jaynes' principle implies that
the best probability assignement $f_{M}(x)$ is the solution of the following
minimization problem:

\begin{equation}
\min KL(f,f_{0})=\min \int_{D}f(x)\ln {\frac{{f(x)}}{{f_{0}(x)}}}dx,
\label{1.1}
\end{equation}
subject to the satisfaction of the following requirements:

i) $f(x) \geq 0, \,\,\, \forall x \in D$;

ii) $\int_D f(x) \, dx=1$;

iii) $I_k(f(x))=0, \,\,\, k=1,2,\dots, M$;

\noindent where $f_{0}(x)$ is the ''prior distribution'' of $X$ and $%
\{I_{k}(f(x))=0,\,\,\,k=1,2,\dots ,M\}$ is a set of relations representing
the information available on the distribution whose $f(x)$ is the density.
In other words Jaynes' prescription is to take the best probability
assignement $f_{M}(x)$ as close as possible to the prior distribution $%
f_{0}(x)$ without however contraddicting the available physical information
as summarized by the constraints $I_{k}$ and the general requirements of any
legitimate density function. Usually, 
\begin{equation}
I_{k}(f(x))=\mu _{k}-\int_{D}x^{k}\,f(x)\ dx,\,\,\,k=1,2,\dots ,M,
\label{1.2}
\end{equation}

where $\mu _{k}$ represents the $k$-th integral moment of the population
having $f(x)$ as {\it pdf}. If the population moments are unknown, it is
possible to replace them with their sample counterparts\cite{baker}. But, it
should be clear that integral moments are not the unique choice. In fact,
when the underlying random variable takes positive values, Novi Inverardi
and Tagliani\cite{novi-tagliani} proposed the use of fractional moments 
\[
{\tilde{\mu}}_{\alpha _{k}}=:E(X^{\alpha _{k}})=\int_{D}x^{\alpha
_{k}}f(x)dx,\,\,\,\alpha _{k}\in I\!\!R,\,\,\,k=0,1,2,\dots ,M,\,\,\,{\tilde{%
\mu}}_{0}=1, 
\]
to represent the available information in the set of constraints given in
Eq. (\ref{1.2}) to spend for recovering the unknown {\it pdf}. With this
setup, the solution of (\ref{1.1}) which gives back the Jaynes' MEp model, 
\begin{equation}
f_{M}(x;\alpha _{k},\lambda _{k})=\exp \{-\sum_{k=0}^{M}\lambda
_{k}\,x^{\alpha _{k}}\}.  \label{1.3}
\end{equation}
The parameter $M$, unknown when the available information consists only in a
sample, represents the order of the model given by the Jaynes' $MEp$ and the 
$\lambda _{k},\,\,\,k=1,2,\dots ,M,$ are the Lagrangian multipliers
associated with the physical constraints $I_{k}(f(x))$.

The main reason that asks for the choice of fractional moments rests on the
fact that integral moments could be very poor tool to extract information
from a sample when the corresponding distribution exhibits fat tails or the
characterizing moments are not integral. In the last case, giving the
fractional moments a better approximation of the characterizing moments, the
performance of the reconstruction density procedure based on them is
expected to be reasonably better than that based on integral moments.

When the only information available consists in a sample, the Jaynes' MEp
needs to be combined with the Akaike selection approach to obtain a complete
procedure for the reconstruction of the underlying unknown {\it pdf}: in
fact Jaynes' $\ $MEp produces an infinite hierarchy of ME models and
Akaike's approach permits to select the optimal member from the hierarchy of
models given by MEp.

It is clear from Eq. (\ref{1.3}) that when constraints involve fractional
moments there is an additional problem to solve: being the exponents $\alpha
_{k}$ of fractional moments new variables to take into account, it needs to
decide not only {\it how many} but also {\it what} fractional moments to
choose in such a way that the estimated density reflects properly the
information contained in a given sample about the unknown probability
distribution. Both of these choices rest on the exploiting of differential
entropy contribution or in other terms choose the $M$ $\alpha $'s exponents
and the $M$ $\lambda $'s coefficients which minimize the $KL$ distance
between $f(x)$ and $f_{M}(x)$; it means the solution of the following
optimization problem: 
\begin{equation}
\min_{M}\left\{ \min_{\alpha }\left\{ \min_{\lambda }\left\{ -{\frac{1}{n}}%
\sum_{i=1}^{n}\ln \left( f_{M}(x_{i};\lambda ,\alpha )\right) +{\frac{M}{n}}%
\right\} \right\} \right\} ,  \label{1.4}
\end{equation}
where $-{\frac{1}{n}}\sum_{i=1}^{n}\ln \left( f_{M}(x_{i};\lambda ,\alpha
)\right) +{\frac{M}{n}}$ represents the sample differential $M$-order model
entropy. The term $M/n$ is proportional to the model order $M$, i.e. to the
number of parameters which we try to estimate using a given sample, and
inversely proportional to the size $n$ of the sample and can be interpreted
in the Akaike's philosophy as a ''penalty term'' which prevents us from
establishing ''too elaborate'' models which cannot be justified by the given
data. Consequently, the parsimony principle becomes an important criterion
whereby we attempt to retain only relevant and useful information and
discard the redundant part of it. More details on the estimation procedure
can be found in Novi Inverardi and Tagliani\cite{novi-tagliani}.

The above technique is applied here to recover from AE's values the analytic
form of the RMS's {\it pdf }that are solution of Eq. (\ref{1.4}) and that
represent a well-tailored model for experimental data distributions. Fig.7
shows, for the three levels of tool wear previously considered, the results
of RMS values {\it pdf} recovering by the ME technique, using only 5
fractional moments. Curves are compared with the experimental histograms
showing a good agreement, especially for newer tools curves, and the visual
inspection of entropy values related to the approximating functions
indicates it decreases whit increasing tool wear level, this representing a
possible further indicator for the phenomena evolution.

\smallskip

{\bf CONCLUSIONS.}

\smallskip Various ways of analysing the basic statistical properties of AE
signals in a TCM application have been illustrated, in which machining
conditions were held fixed throughout the experiment, in order to limit the
number of varying parameters that might affect the behaviour of the observed
quantities. The analysis has been performed on signals gathered at the same
time using two different AE sensors, enabling a comparative analysis in
which, for both sensors, some interesting features, till now not
sufficiently underlined, have emerged. In particular, both AE time series
and their associated RMS values experimental frequency distributions have
been derived, allowing to analyse how tool wear affects such statistical
features in the kind of situations investigated in our experiment. For what
concerns the RMS values, the shape of the curves indicates a noticeable
shift towards lower levels of the modal value for increasing levels of wear,
this indicating a reduced AE activity, together to a reduction in the signal
variability and a change in the skewness towards values compatible with a
symmetrical shape.

A Beta function model for describing the RMS's {\it pdf} has been tested,
and the residuals in the best-fitted function indicate that a more
appropriate fitting model should be sought. A much better agreement has been
reached by resorting to a ME technique by means of which the general
Hausdorff moment problem has been tackled in an original way by using only
few sampling fractional moments, this providing a better tailored analytic
form for the RMS's experimental distributions than previously proposed
models. It has been also observed that the entropy of the functions
monotonically changes for wear increasing. On the other hand, the physical
meaning of the Lagrange multipliers $\lambda _{j}$ obtained in this fitting
function reconstruction, (or the equivalent fractional moments order $\alpha
_{j}$) is not clear, and future efforts should be done to clarify this
aspect.

Particularly interesting are the statistical properties of the time series,
in which power laws in the frequency distributions have been identified, in
accordance with what has been pointed as a feature of acoustic emission
phenomena in numerous other fields. In particular, the evidence of the
non-gaussianity of the process would make it reasonable to tackle the signal
blind deconvolution problem by means of higher order statistics (HOS)\cite
{cadzow}. The recovering, only from the observed output, of the unknown
original signal before it had been altered by the sensor response and the
measurement chain, would be a fundamental step towards a deeper
understanding of AE phenomena associated to TCM and more general
applications as well.


\begin{references}
\bibitem{footnote}  Fitting the function given in Eq.1 requires a
non-trivial effort, as this function cannot be linearised, so that the
fitting process requires initial seed values; these might be obtained using
the r,s estimation technique described above. Additionally, fully automating
the fitting process requires appropriate handling of specific exceptions
which can occur in non-linear regressions (e.g. singular matrix conditions,
reaching iteration limits without having satisfied tolerance criteria,
etc.). In this paper no attempt at automating this process has been made.
\end{references}
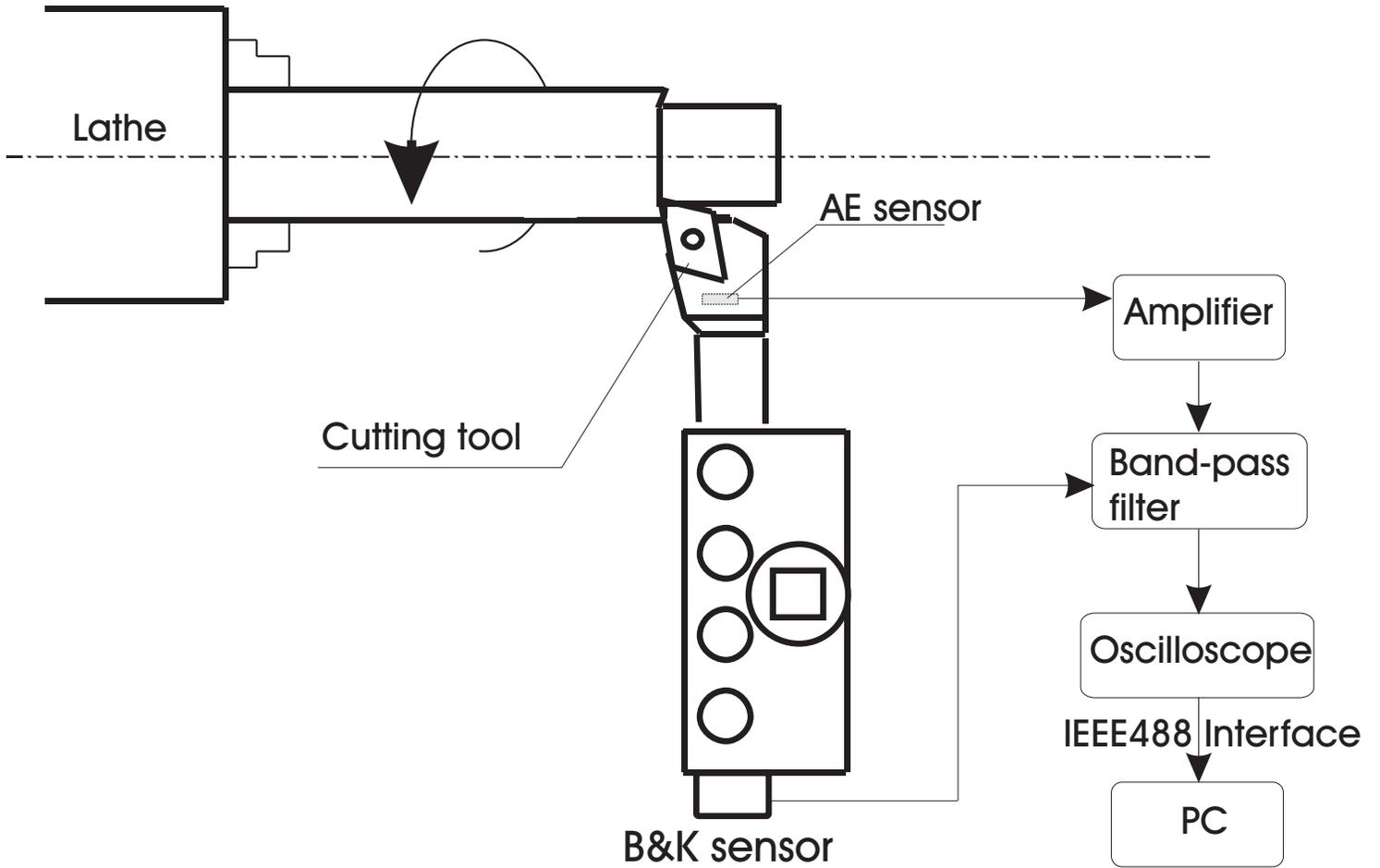
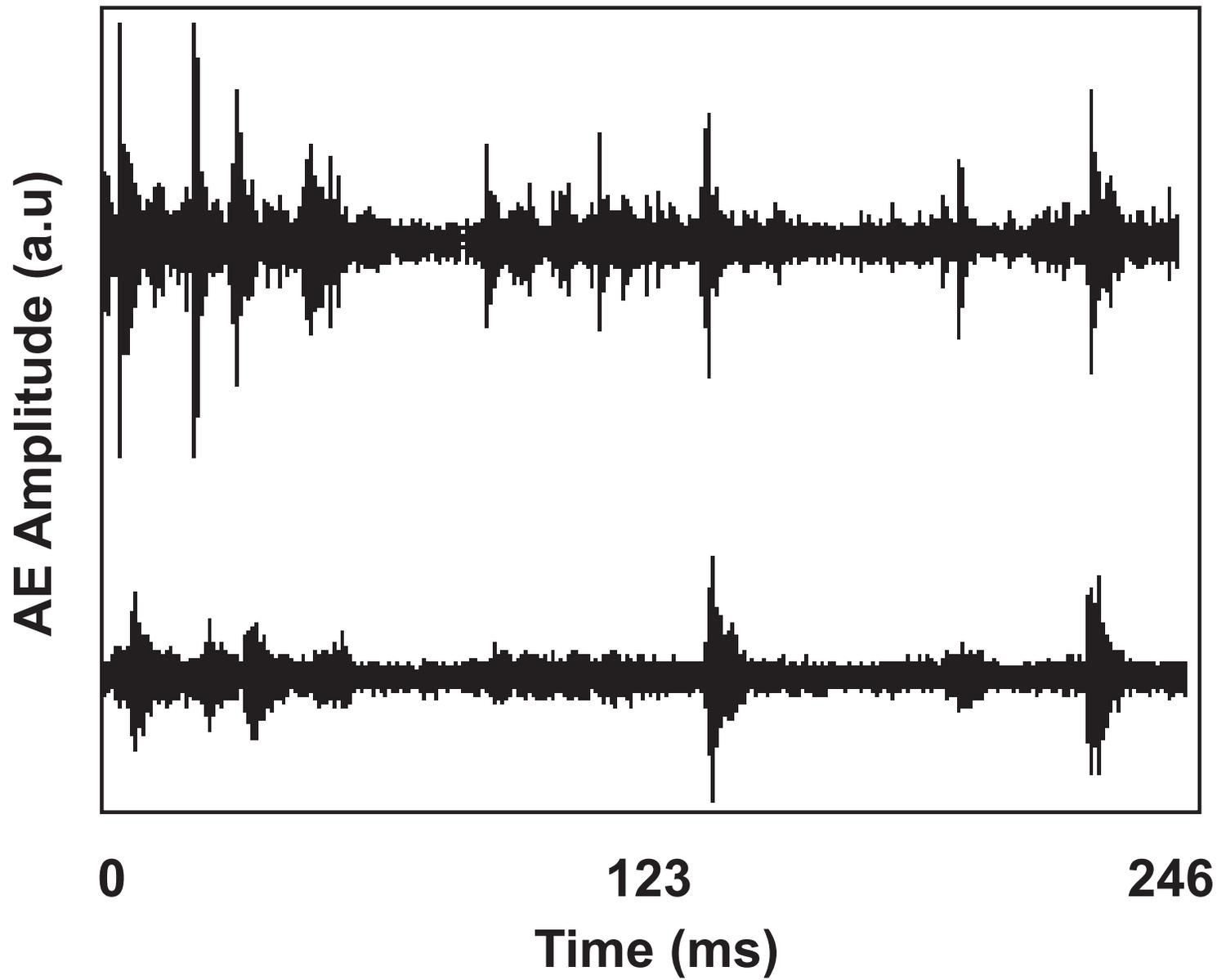
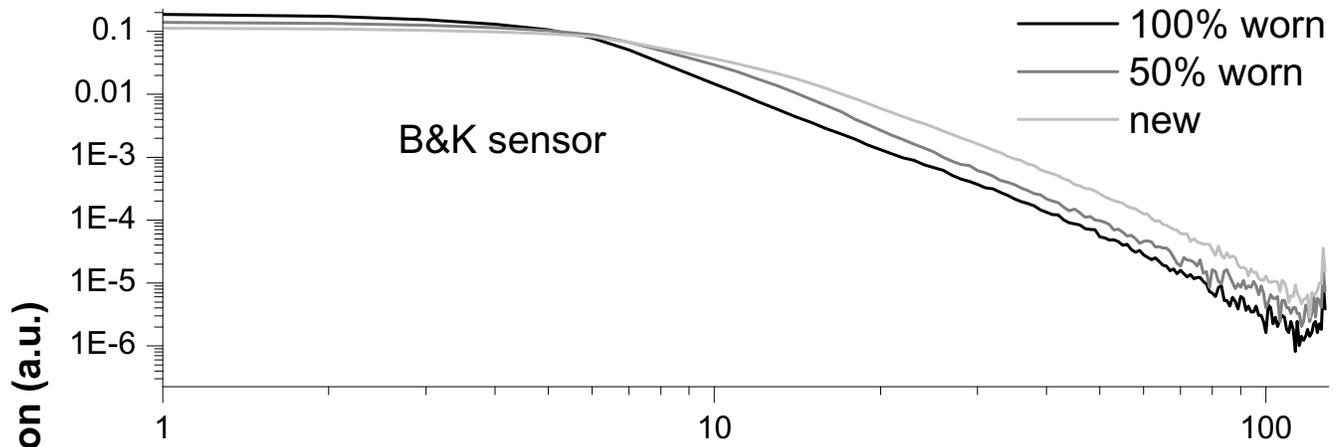
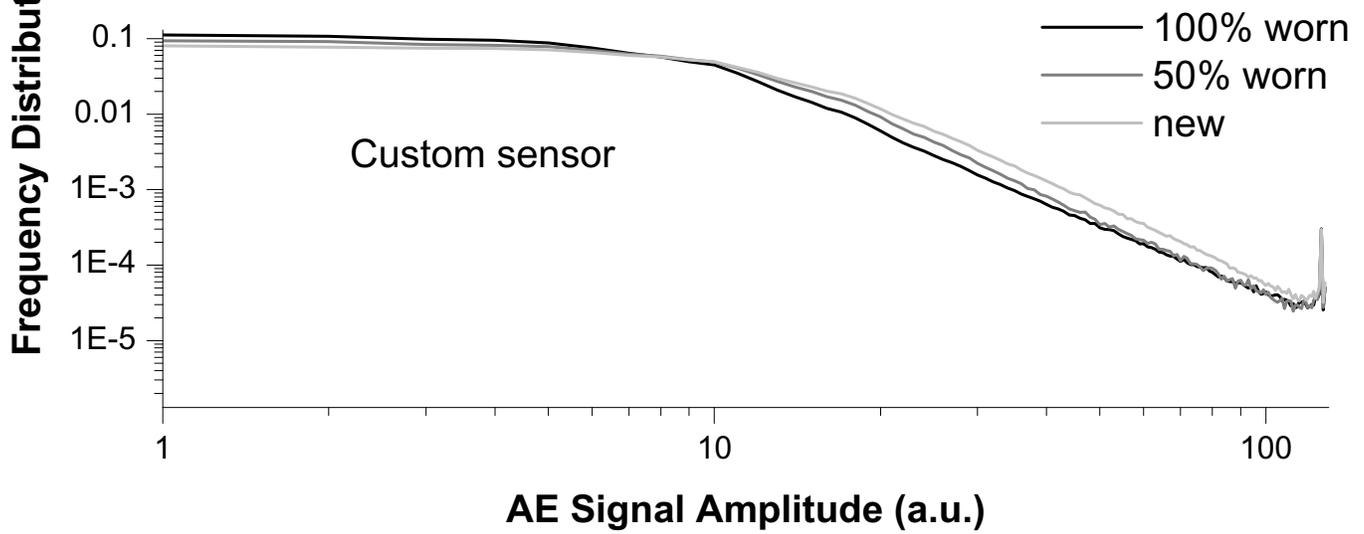
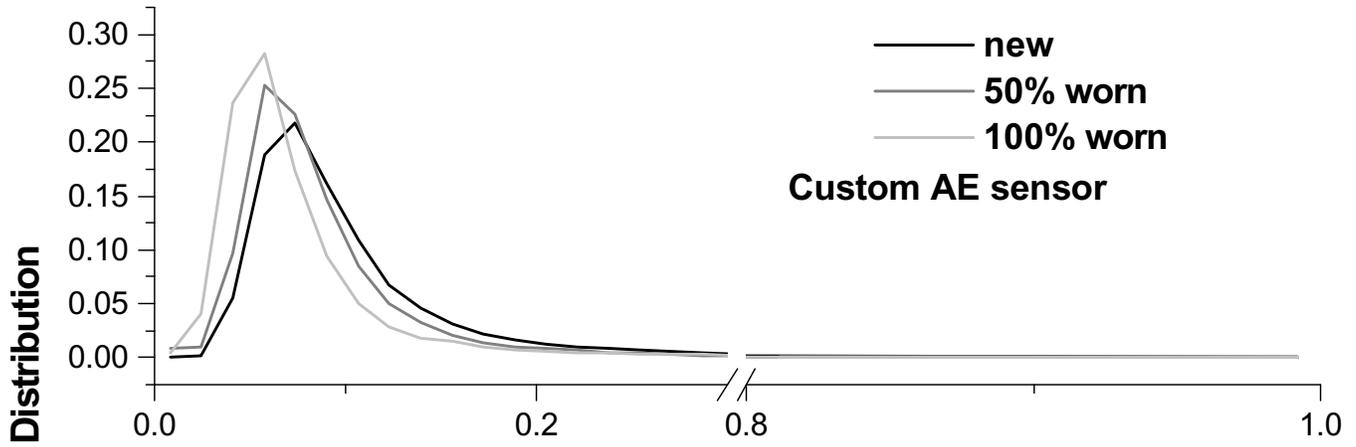
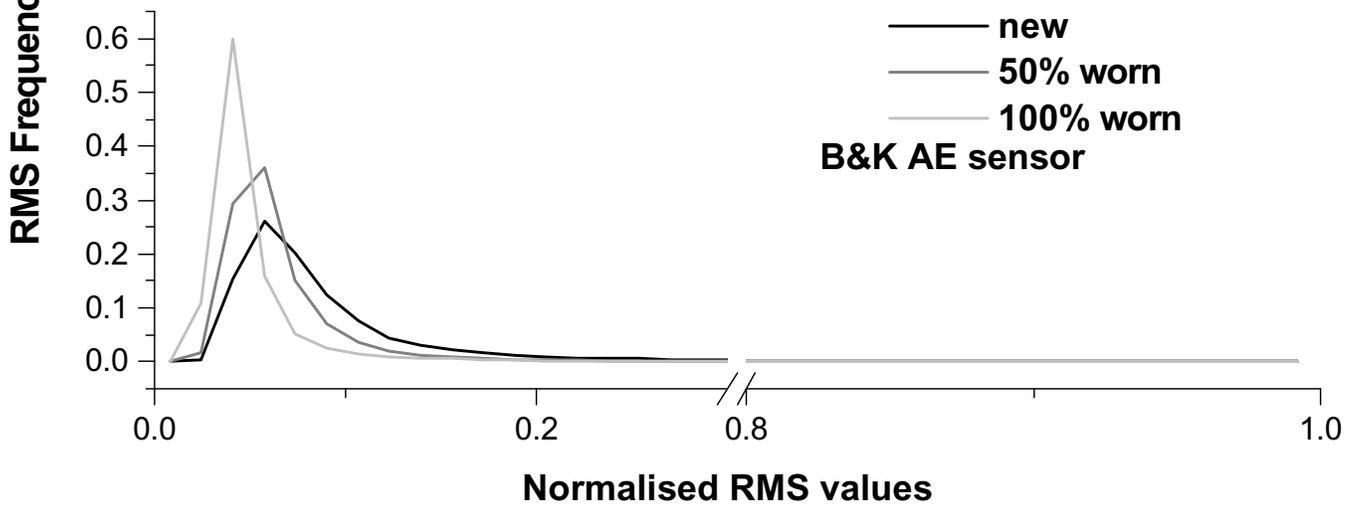
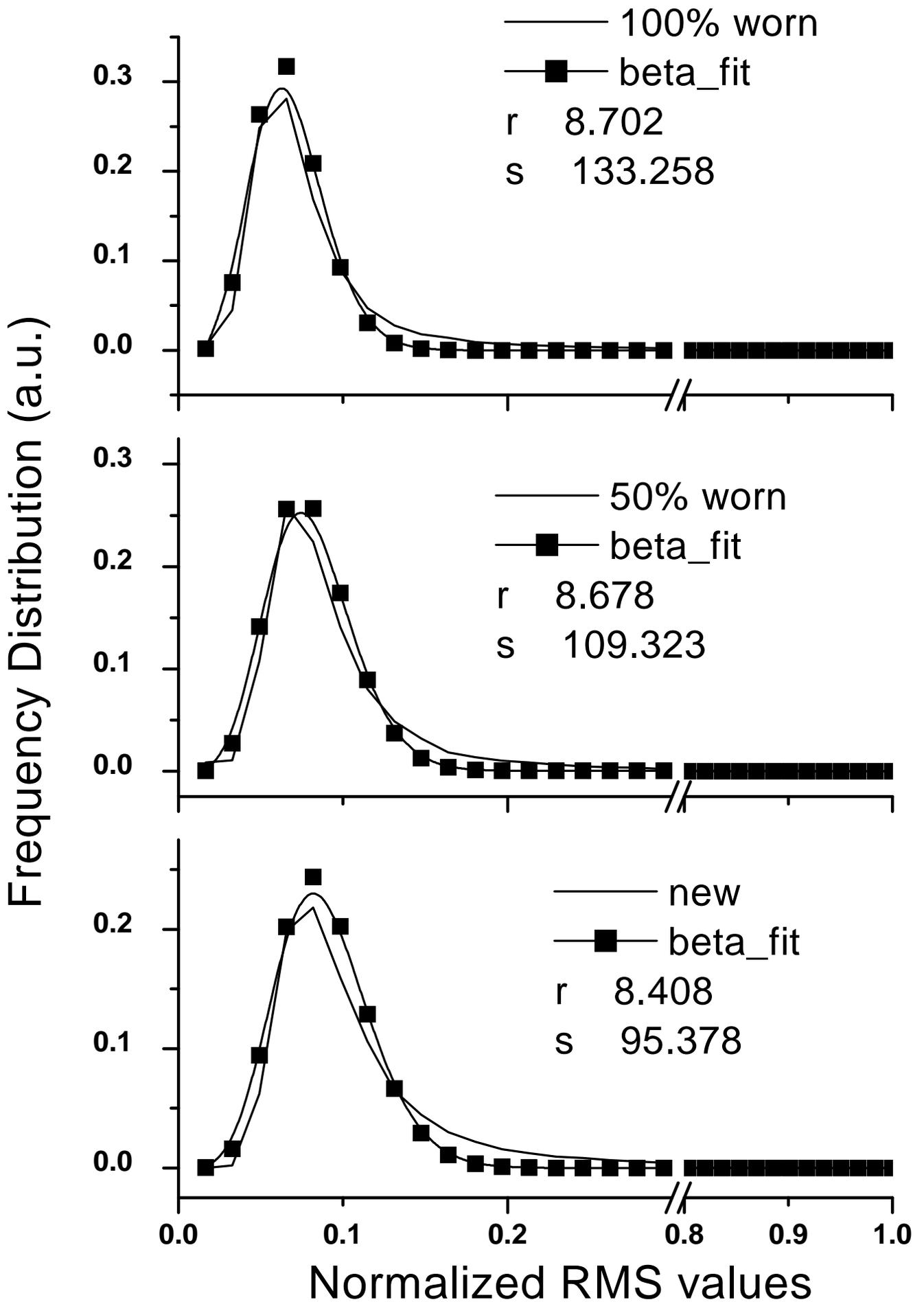
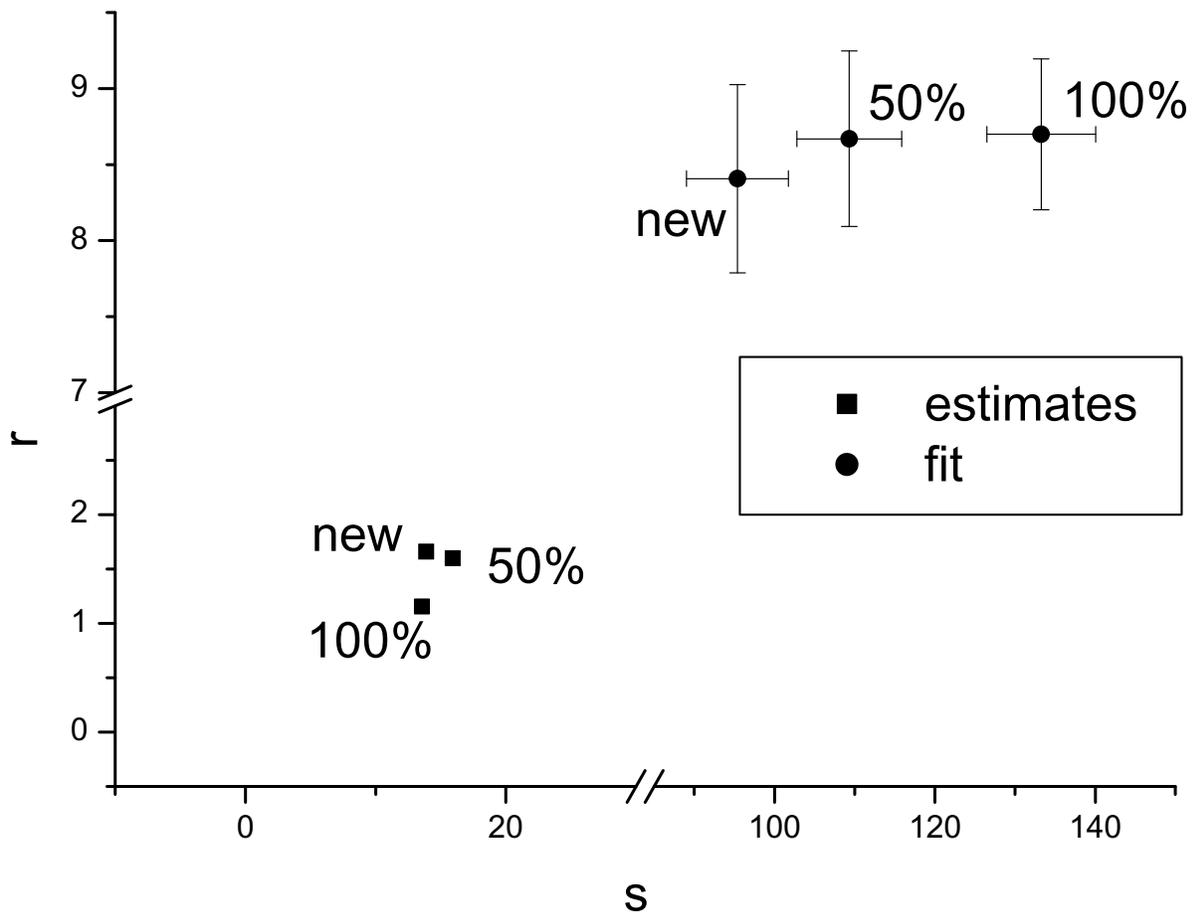
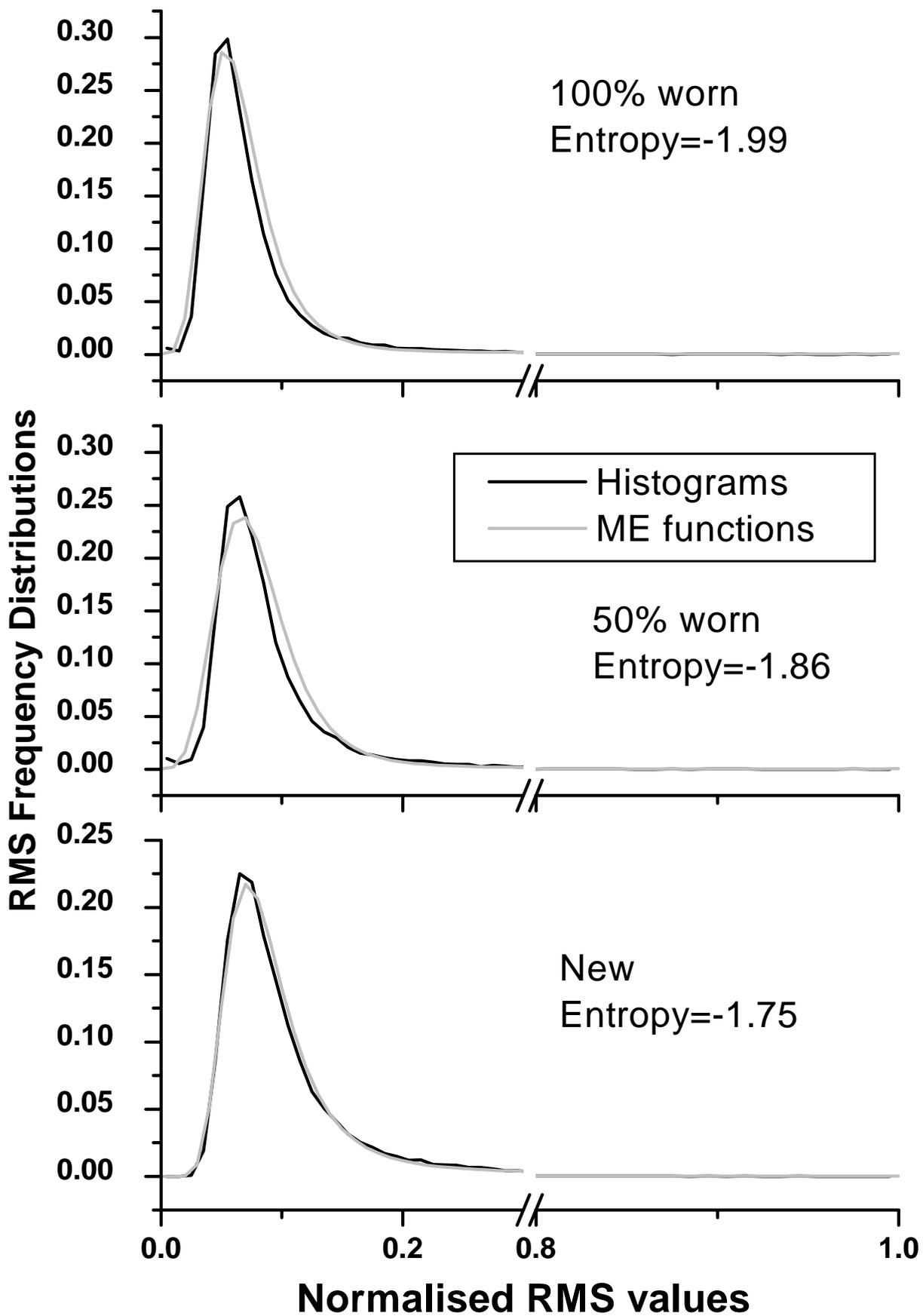
\end{document}